\documentstyle[12pt,epsfig]{article}
\newcommand{\be}{\begin{equation}}
\newcommand{\ee}{\end{equation}}
\newcommand{\beq}{\begin{equation}}
\newcommand{\eeq}{\end{equation}}
\newcommand{\ba}{\begin{eqnarray}}
\newcommand{\ea}{\end{eqnarray}}
\newcommand{\bea}{\begin{eqnarray}}
\newcommand{\eea}{\end{eqnarray}}

\begin{document}
\baselineskip=15.5pt \pagestyle{plain} \setcounter{page}{1}


\def\del{{\partial}}
\def\vev#1{\left\langle #1 \right\rangle}
\def\cn{{\cal N}}
\def\co{{\cal O}}
\newfont{\Bbb}{msbm10 scaled 1200}     
\newcommand{\mathbb}[1]{\mbox{\Bbb #1}}
\def\IC{{\mathbb C}}
\def\IR{{\mathbb R}}
\def\IZ{{\mathbb Z}}
\def\RP{{\bf RP}}
\def\CP{{\bf CP}}
\def\Poincare{{Poincar\'e }}
\def\tr{{\rm tr}}
\def\tp{{\tilde \Phi}}

\def\TL{\hfil$\displaystyle{##}$}
\def\TR{$\displaystyle{{}##}$\hfil}
\def\TC{\hfil$\displaystyle{##}$\hfil}
\def\TT{\hbox{##}}
\def\HLINE{\noalign{\vskip1\jot}\hline\noalign{\vskip1\jot}} 
\def\seqalign#1#2{\vcenter{\openup1\jot
  \halign{\strut #1\cr #2 \cr}}}
\def\lbldef#1#2{\expandafter\gdef\csname #1\endcsname {#2}}
\def\eqn#1#2{\lbldef{#1}{(\ref{#1})}%
\begin{equation} #2 \label{#1} \end{equation}}
\def\eqalign#1{\vcenter{\openup1\jot
    \halign{\strut\span\TL & \span\TR\cr #1 \cr
   }}}
\def\eno#1{(\ref{#1})}
\def\href#1#2{#2}
\def\half{{1 \over 2}}

\def\ads{{\it AdS}}
\def\adsp{{\it AdS}$_{p+2}$}
\def\cft{{\it CFT}}

\newcommand{\ber}{\begin{eqnarray}}
\newcommand{\eer}{\end{eqnarray}}

\newcommand{\beqar}{\begin{eqnarray}}
\newcommand{\cN}{{\cal N}}
\newcommand{\cO}{{\cal O}}
\newcommand{\cA}{{\cal A}}
\newcommand{\cT}{{\cal T}}
\newcommand{\cF}{{\cal F}}
\newcommand{\cC}{{\cal C}}
\newcommand{\cR}{{\cal R}}
\newcommand{\cW}{{\cal W}}
\newcommand{\eeqar}{\end{eqnarray}}
\newcommand{\th}{\theta}
\newcommand{\lm}{\lambda}\newcommand{\Lm}{\Lambda}
\newcommand{\eps}{\epsilon}
\newcommand{\pa}{\paragraph}
\newcommand{\pt}{\partial}
\newcommand{\de}{\delta}
\newcommand{\De}{\Delta}
\newcommand{\lb}{\label}


\newcommand{\nonu}{\nonumber}
\newcommand{\oh}{\displaystyle{\frac{1}{2}}}
\newcommand{\dsl}
  {\kern.06em\hbox{\raise.15ex\hbox{$/$}\kern-.56em\hbox{$\partial$}}}
\newcommand{\id}{i\!\!\not\!\partial}
\newcommand{\as}{\not\!\! A}
\newcommand{\ps}{\not\! p}
\newcommand{\ks}{\not\! k}
\newcommand{\D}{{\cal{D}}}
\newcommand{\dv}{d^2x}
\newcommand{\Z}{{\cal Z}}
\newcommand{\N}{{\cal N}}
\newcommand{\Dsl}{\not\!\! D}
\newcommand{\Bsl}{\not\!\! B}
\newcommand{\Psl}{\not\!\! P}
\newcommand{\eeqarr}{\end{eqnarray}}
\newcommand{\ZZ}{{\rm \kern 0.275em Z \kern -0.92em Z}\;}

\begin{titlepage}

\leftline{OUTP-04-14-P} \leftline{\tt hep-th/0407152}

\vskip -.8cm


\begin{center}

\vskip 1.5 cm

{\LARGE Glueballs, symmetry breaking and axionic strings in  
non-supersymmetric deformations of the Klebanov-Strassler
background} \vskip .3cm



\vskip 1.cm

{\large Martin Schvellinger{\footnote{ martin@thphys.ox.ac.uk}}}

\vskip 0.6cm

{\it The Rudolf Peierls Centre for Theoretical Physics, \\
Department of Physics, University of Oxford. \\ 1 Keble Road,
Oxford, OX1 3NP, UK.}


\vspace{1.7cm}

{\bf Abstract}

\end{center}

We obtain an analytic solution for a pseudo-scalar massless perturbation of a 
non-su\-per\-symme\-tric deformation 
of the warped deformed conifold. This allows us to study D-strings in the 
infrared limit of non-supersymmetric deformations of the Klebanov-Strassler 
background. They are interpreted as axionic strings in the dual field theory. 
Following the arguments of hep-th/0405282, the axion is a massless 
pseudo-scalar glueball which is present in the supergravity fluctuation
spectrum and it is interpreted as the Goldstone boson of the
spontaneously broken $U(1)_B$ baryon number symmetry, being the gauge theory
on the baryonic branch. 

\noindent

\end{titlepage}

\newpage

\tableofcontents

\newpage


\vfill

\section{Introduction} 

After the discovery of the AdS/CFT duality \cite{Maldacena:1997re} (see also 
\cite{adscft}) the natural expectation has been to construct 
supergravity dual versions of the large $N$ limit of minimally supersymmetric 
Yang-Mills theories, as well as non supersymmetric ones. 
In the quest of the holographic dual description of a 
realistic field theory the findings of the supergravity duals of  
${\cal {N}}=1$ SYM theories \cite{Klebanov:2000hb,Maldacena:2000yy} have 
certainly been remarkable concrete realizations of this idea. Although in 
the UV these dual models do not behave like a conventional asymptotically 
free gauge theory, they are supergravity duals of certain confining 
${\cal {N}}=1$ SYM theories. This has led to expect that these two 
models would fall in the same infrared universality class as the large $N$ 
limit of the pure ${\cal {N}}=1$ super-gluodynamics. However, as it was very 
recently shown by Gubser, Herzog and Klebanov \cite{Gubser:2004qj}, it turns 
out that at least for the Klebanov-Strassler (KS) solution this is not quite the case.
In that paper the authors have reached this conclusion by studying a certain 
perturbation of the RR 2-form field which also mixes with the RR 4-form field\footnote{Since 
this particular supergravity perturbation is interpreted as an axion in its 
dual field theory description, we will call it 
axionic perturbation even when this is given by a certain combination of perturbations of the 
RR 2-form and 4-form fields, and it does not include the axion of 
ten dimensional type IIB supergravity.}. They interpreted  
D-strings at the bottom of the warped deformed conifold as axionic strings
in the dual $SU(N+M) \times SU(N)$ ${\cal {N}}=1$ SYM theory, finding that the
axion is a massless pseudo-scalar glueball present in the supergravity fluctuation 
spectrum. This was interpreted as the Goldstone boson of spontaneously broken $U(1)_B$
baryon number symmetry. This result provides further evidence for an earlier 
conjecture that the field theory is on the baryonic branch \cite{Aharony:2000pp}.

On the other hand, non-supersymmetric deformations of the Klebanov-Strassler 
background have been studied in 
\cite{Borokhov:2002fm,Apreda:2003gc,Bigazzi:2003ui,Apreda:2003gs,Kuperstein:2003yt, 
Kuperstein:2003ih,Bigazzi:2004yt}. All these non-supersymmetric deformations do not
include any perturbation similar to the one discussed in \cite{Gubser:2004qj}.  
Hence, it is interesting to investigate the possibility of
including axionic perturbations on non-supersymmetric deformations of the Klebanov-Strassler 
background. Indeed, the motivation of this paper is to obtain an analytic solution 
for an axionic non-super\-symme\-tric deformation of the KS background, and
on this scenario, to show how to extend the results of Gubser, Herzog and 
Klebanov to non-supersymmetric deformations. We will consider the analytic solution obtained
by Kuperstein and Sonnenschein \cite{Kuperstein:2003yt} which is a non-singular, non-supersymmetric
first order deformation of the KS solution, preserving the $SU(2) \times SU(2)$ global
symmetry of the KS background. The only non-vanishing vacuum expectation values (VEVs) are those
corresponding to operators invariant under this symmetry, i.e. the baryonic operators.
Thus, it seems that the theory is on the baryonic branch \cite{Aharony:2000pp}. 
Moreover, it would be expected that
certain non-supersymmetric deformations of the original background developed some
similar properties associated with axionic perturbations as the KS solution itself. 
In this way, we want to show that in this
non-supersymmetric case it is possible to find a massless pseudo-scalar glueball which is
present in the supergravity fluctuation spectrum. We present an explicit  
non-supersymmetric example that provides new evidence in favor of the hypothesis 
that the $U(1)_B$ baryon number symmetry is broken by expectation values of 
baryonic operators. This result agrees with the corresponding one in the case of the 
supersymmetric KS background reported in \cite{Gubser:2004qj}. However, there are
certain differences due to the fact that superymmetry is not preserved. 
Interestingly, in breaking supersymmetry there appears a $(0,3)$ form which is related 
to a non-normalizable mode.  Moreover, in the
non-supersymmetric deformation of Kuperstein and Sonnenschein \cite{Kuperstein:2003yt}
there is a second non-normalizable mode related to the deformation of the metric. 
Remarkably, the supergravity mode associated with the axion that we found 
turns out to be normalizable, which for this non-supersymmetric deformation allows us to
perform a parallel analysis to the one done in \cite{Gubser:2004qj} for the 
supersymmetric warped deformed conifold.  We will discuss the implications of this 
axionic perturbation on the non-supersymmetric solution in the dual field theory description.

The paper is organized as follows.
We firstly review the main idea of \cite{Gubser:2004qj} for the supersymmetric 
KS solution. This is done in section 2. Then, in section 3 we explicitly show how to 
extend the axionic perturbation ansatz for the analytic non-supersymmetric 
deformations of the Klebanov-Strassler solution obtained by Kuperstein and Sonnenschein 
\cite{Kuperstein:2003yt}. 
We explicitly solve the linearized equations for the perturbation ansatz
obtaining the general solution. We are able to choose the appropriate boundary conditions
to obtain an IR and UV well-behaved solution for the supergravity fluctuation.
The non-supersymmetric deformation is controlled by a 
parameter such that the background becomes the KS solution once the
deformation vanishes. In this perturbative approach the axionic perturbation of the 
non-supersymmetric deformation leads to a normalizable supergravity mode.
In section 4 we study some properties of the dual field theories, both supersymmetric 
and non-supersymmetric ones, when considering the axionic perturbation.
Furthermore, we very briefly discuss about the Pando Zayas-Tseytlin background \cite{PandoZayas:2000sq} 
where it was conjectured that the dual field theory would be on the mesonic branch 
\cite{Aharony:2000pp}. In section 5 we discuss our results, as well as related 
open questions that we consider interesting to investigate further.

~

\section{Axionic strings in the KS background}

We would like to study the effects of an axionic string in 
non-supersymmetric deformations of the KS background,
and compare them with the case studied in \cite{Gubser:2004qj} for the
supersymmetric warped deformed conifold. In order to do this, 
in this section we review some of the results of \cite{Gubser:2004qj} which
will be relevant for our purposes. Consider a 
D1-brane extended in two of the four dimensions of $R^{3,1}$. The D1-brane
carries electric charge under the R-R three-form field strength $F_3$. Hence,
one can think that an axion $a$ in four dimensions defined so that
\bea
*_4da &=& \delta F_3 \, ,
\eea
experiences monodromy when one makes a loop around the D1-brane.
The symbol $*_4$ represents the Hodge dual operation on the 4d Minkowski space-time,
while $*$ labels the Hodge dual operation on the 10d background.
Note that a very simple ansatz for the axion is $a=a(t) = f_1 \, t$, leading to 
the following ansatz for the perturbation of the RR three-form field strength 
$F_3$
\bea
\delta F_3
 &=& *_4da = f_1 \, dx^1 \, \wedge \, dx^2 \, \wedge \, dx^3 \, .
\eea
Thus, given the above ansatz for the perturbation, one should add whatever terms 
are necessary in order to solve the linearized equations of motion. Such a solution 
would represent a zero-momentum axion. For the warped deformed conifold we will use 
the notation given in references \cite{Herzog:2001xk,Herzog:2002ih} (see Appendix B).

The proposed perturbation ansatz for the KS background given in reference \cite{Gubser:2004qj} is
\bea
\delta H_3 &=& 0 \, , \nonumber \\
\delta F_3 &=& f_1 \, *_4 da + f_2^0(\tau)  \, da \, \wedge \, dg^5 + 
               {\dot f_2^0}(\tau)  \, da \, \wedge \, d\tau \, \wedge \, g^5 \, , \nonumber \\
\delta F_5 &=& (1 + *) \, \delta F_3
 \, \wedge \, B_2 \, . \label{dF50}
\eea
We assume that the perturbations of the rest of the fields are zero. Dot denotes
derivative with respect to $\tau$. The axion is assumed to be a function of
the $(t, x^1, x^2, x^3)$ coordinates, therefore 
$d *_4 da$ vanishes. The sum of the second two terms in the above 
ansatz for $\delta F_3$ leads to an exact two form 
$-d(f_2^0 \, da \, \wedge \, g^5)$. We must check that the above perturbation 
ansatz satisfies the following EOM derived from type IIB supergravity action
(see Appendix A)
\bea
d(* F_3) &=& g_s \, F_5 \, \wedge \, H_3 \, , \label{eom-F31}
\\
d(* H_3) &=& -  g_s \, F_5 \, \wedge \, F_3  \, ,  \label{eom-h31} 
\eea
together with the Bianchi identities Eqs.(\ref{bianchi0}), (\ref{bianchi1}) and (\ref{bianchi2}).
The first of the Bianchi equations implies that $d\delta F_3 = 0$, and therefore
$f_1$ is nothing but a constant that is set to 1 \cite{Gubser:2004qj}.
On the other hand, from the second Bianchi identity  Eq.(\ref{bianchi1}) we have 
$d(\delta F_5) = H_3 \, \wedge \, \delta F_3$, which is
satisfied for harmonic $a$. One can show this using the identities (\ref{id1}) and (\ref{id2}) 
given in Appendix B. 

In order to check whether the perturbation ansatz satisfies Eq.(\ref{eom-F31}), one
first obtains the Hodge dual of $\delta F_3$ which is given by
\bea
*\delta F_3 &=& f_1 \, h_0^2(\tau) \, \sinh^2\tau \, \frac{\epsilon^4}{96} \,
da \, \wedge \, d\tau \, \wedge \, g^1 \, \wedge \, g^2 \, \wedge \, g^3  \, 
\wedge \, g^4 \, \wedge \, g^5
\nonumber \\
& & - \frac{f_2^0(\tau) \, \epsilon^{4/3}}{6 \, K^2(\tau)}  \, (*_4 da)  \, \wedge \, d\tau 
 \, \wedge \, g^5  \, \wedge \, dg^5 \nonumber \\
& & + \frac{3}{8} \, {\dot f_2^0}(\tau) \, \epsilon^{4/3} \, K^4(\tau) \, \sinh^2\tau \, \, 
(*_4 da) \, \wedge \, g^1 \, \wedge \, g^2 \, \wedge \, g^3  \, \wedge \, g^4 \, . \label{F30}
\eea
Therefore
\bea
d(*\delta F_3) &=& \frac{3 \, \epsilon^{4/3}}{4} \, 
\left( -\frac{d}{d\tau} [{\dot f_2^0}(\tau) \, K^4 \, \sinh^2\tau ] 
+ \frac{8 \, f_2^0}{9 \, K^2}\right) 
\, (*_4 da)  \, \wedge \, d\tau  \, \wedge \, \omega_2  \, \wedge \, \omega_2 \, ,
\eea
while
\bea
\delta F_5 \, \wedge \, H_3 &=& 2 \, \frac{f_1}{4} \, (g_s M \alpha')^2   \,
\frac{d(k(\tau) \, f(\tau))}{d\tau} 
\, (*_4 da)  \, \wedge \, d\tau  \, \wedge \, \omega_2  \, \wedge \, \omega_2 \, ,
\eea
where identities (\ref{id}), (\ref{idi}) and Eq.(\ref{derivativefk}) have been used.
Definitions of functions $K(\tau)$, $F(\tau)$, $f(\tau)$ and $k(\tau)$ are given in Appendix B.
With these expressions,  from Eq.(\ref{eom-F31}) one gets the following second order ordinary
differential equation
\bea
-\frac{d}{d\tau} [{\dot f_2^0}(\tau) \, K^4 \, \sinh^2\tau ] + \frac{8 \, f_2^0}{9 \, K^2} &=& 
\frac{(g_s  M \alpha')^2}{3 \, \epsilon^{4/3}} \, (\tau \, \coth\tau-1) \, 
\left( \coth\tau-\frac{\tau}{\sinh^2\tau}\right) \label{ODEf20}
\, .
\eea
Now, in order to solve the above ODE one starts with the homogeneous differential equation
\bea
-\frac{d}{d\tau} [{\dot f_2^0}(\tau) \, K^4 \, \sinh^2\tau ] + \frac{8 \, f_2^0}{9 \, K^2} &=& 
0 \, ,
\eea
which is solved by the functions
\bea
f_2^{0 \, (1)}(\tau) &=& [\sinh(2 \tau) - 2 \tau]^{1/3} \, , \\
f_2^{0 \, (2)}(\tau) &=& [\sinh(2 \tau) - 2 \tau]^{-2/3} \, .
\eea
The general solution of an inhomogeneous equation of the form
\bea
\frac{d^2 y(\tau)}{d\tau^2} + p(\tau) \, \frac{d y(\tau)}{d\tau} 
+ q(\tau) \, y(\tau) &=& j(\tau) \, ,
\eea
is 
\bea
y(\tau) &=& c_1 \, y_1(\tau) + c_2 \, y_2(\tau) + y_p(\tau) \, ,
\eea
where $c_1$ and $c_2$ are constants, while $y_1$ and $y_2$ are independent solutions
of the homogeneous ODE ($f_2^{0 \, (1)}$ and $f_2^{0 \, (2)}$, respectively), and $y_p(\tau)$ 
is a particular solution of the inhomogeneous equation given by
\bea
y_p(\tau) &=& -y_1(\tau) \, \int_0^\tau \, dx \, \frac{y_2(x)}{W(y_1, y_2)(x)} \, j(x) +
y_2(\tau) \, \int_0^\tau \, dx \, \frac{y_1(x)}{W(y_1, y_2)(x)} \, j(x) \, ,
\eea
where the Wronskian is defined as usual 
\bea
W(y_1, y_2)(x) &=& y_1(x) \, y'_2(x) - y_2(x) \, y'_1(x) \, , \label{wronskian}
\eea
and prime denotes derivative with respect to $x$.
In the present case it is not difficult to obtain the solution
\bea
f_2^0(\tau) &=& - 2 \, \frac{\epsilon^{4/3}}{12} \, \frac{1}{K^2(\tau) \, \sinh^2\tau} \,
\int_0^\tau \, dx \, h_0(x) \, \sinh^2x \, . \label{f20}
\eea
Indeed, this solution $f_2^0(\tau)$ behaves like $\tau$ for 
small $\tau$, while it falls as $\tau \, e^{-2/3 \, \tau}$ for $\tau \rightarrow \infty$.
In section 3 we will give the explicit asymptotic expressions. Notice that we have used
the label $0$ to indicate that $f_2^0(\tau)$ is the function corresponding to
the supergravity fluctuation on the supersymmetric warped deformed conifold. 
This notation will be useful in the next section.
The related definitions and conventions are explicitly given in Appendix B.

Now, an important point is to show that $\delta F_3$ is normalizable. To do so
it is necessary to integrate the fluctuation such that the integral of  
$\sqrt{|g|} \, |\delta F_3|^2$ over $\tau$, where $g$ is the ten dimensional
metric, must be finite. In addition, note that 
\bea
\sqrt{|g|} \, |\delta F_3|^2 \, d^{10}x &=& 
\delta F_3 \, \wedge \, *\delta F_3 \, .
\eea
The fact that the above integral reduces to the sum of 
three single well-behaved integrals at small $\tau$ and falls like $e^{-2\tau/3}$
for $\tau \rightarrow \infty$, guarantees that the perturbation is normalizable. This
is a normalizable zero-mode of the KS background \cite{Gubser:2004qj}. 
We will return to the analysis of this perturbation in the next sections, in order
to compare it with related issues from non-supersymmetric deformations of the KS solution.

~

\section{Axionic strings in non-supersymmetric deformations of the KS background}

In this section, we study axionic strings in non-supersymmetric deformations of 
the KS background. An analytic non-supersymmetric deformation of the warped 
deformed conifold has been obtained by Kuperstein and Sonnenschein \cite{Kuperstein:2003yt}. 
Their solution is based on an expansion of the fields in terms of a parameter ${\tilde \delta}$, 
which accounts for the non-supersymmetric deformation of the background. The explicit 
deformations were obtained using a modification of the superpotential method introduced in
\cite{Borokhov:2002fm}. This will be briefly reviewed in this section in order to
have the necessary background to perform our calculations.

The idea is that the axionic perturbation ansatz can be understood as a perturbation
of the non-supersymmetric deformations of the KS background. 
Therefore, at first order in ${\tilde \delta}$
we will have the same EOM, as well as the first order equations derived from the
superpotential for the deformed non-supersymmetric background
like in \cite{Kuperstein:2003yt}. On the other hand, the perturbation of the 2-form and 4-form fields
will be controlled by perturbation equations analogous to the ones in the previous section.

In order to show explicitly how it works, let us remember that we deal with the
linearized EOM. Therefore, the dilaton EOM leads to the same equation as in the
supersymmetric case, i.e. $e^\phi \, F_3 \, \wedge \, * F_3 = 
e^{-\phi} \, H_3 \, \wedge \, *H_3$. In addition, there is the usual relation $g_s=e^\phi$. 
The equation for the five-form field strength has the form 
$dF_5 = H_3 \, \wedge \, F_3$. All these equations are satisfied 
by the non-supersymmetric deformation of Kuperstein and Sonnenschein, since
they are satisfied by the first order equations for the deformations.
The equations for the perturbation ansatz look similar 
to the corresponding ones in section 2. We will show this for $f_2(\tau)$ in this section.

Analytic non-supersymmetric deformations of the 
KS background have been constructed by Kuperstein and Sonnenschein using the superpotential method. 
This method has been extensively used 
for studying gravitational RG flows in five-dimensional gauged supergravity 
\cite{DeWolfe:1999cp,Skenderis:1999mm,deBoer:1999xf,Borokhov:2002fm,Kuperstein:2003yt}. 
Here we briefly review this method following  \cite{Borokhov:2002fm,Kuperstein:2003yt}. 
We use the notation for the warped deformed conifold metric of \cite{Herzog:2001xk,Herzog:2002ih}, 
while for the non-supersymmetric deformation we follow \cite{Kuperstein:2003yt}. 
Let us consider the most general metric
with $SU(2) \times SU(2)$ isometry, that includes the Klebanov-Strassler solution by setting
$\tilde \delta=0$. We can write the metric using the following ansatz
\bea
ds_{10}^2 &=& 2^{1/2} \, 3^{3/4} \, [ e^{-5 q(\tau) + 2 Y(\tau)} \, (-dt^2+d{\vec {x}}^2)
+ \frac{1}{9} \, e^{3 q(\tau) - 8 p(\tau)} \, (d\tau^2+(g^5)^2) + \nonumber \\
& & \frac{1}{6} \, e^{3 q(\tau) + 2 p(\tau) + y(\tau)} \, ((g^1)^2+(g^2)^2) +
\frac{1}{6} \, e^{3 q(\tau) + 2 p(\tau) - y(\tau)} \, ((g^3)^2+(g^4)^2)  ] \, .
\label{metric0}
\eea
For definitions of the one forms $g^i$'s we refer the reader to Appendix B.
We assume the axion as a function $a=a(t, x^1, x^2, x^3)$, 
while the dilaton is $\phi=\phi(\tau)$. We consider the following ansatz for the fields
\bea
B_2 &=& - ({\tilde {f}}(\tau) \, g^1 \, \wedge \, g^2 + {\tilde {k}}(\tau) \, 
g^3 \, \wedge \, g^4) \, , \\
%
F_3 &=& 2 \, P \, g^5 \, \wedge \, g^3 \, \wedge \, g^4 + d[{\tilde {F}}(\tau) \, 
( g^1 \, \wedge \, g^3 + g^2 \, \wedge \, g^4)]  \, , \\
F_5 &=& {\cal {F}}_5 + *{\cal {F}}_5 \, ,
\eea
where
\bea
{\cal {F}}_5 &=& - {\tilde {L}}(\tau) \, 
g^1 \, \wedge \, g^2 \, \wedge \, g^3 \, \wedge \, g^4 \, \wedge \, g^5 \, , 
\eea
with
\bea
{\tilde {L}}(\tau) &=& Q + {\tilde {f}}(\tau) \, (2 P - {\tilde {F}}(\tau)) 
+ {\tilde {k}}(\tau) \, {\tilde {F}}(\tau) \, ,   
\eea
while the perturbation ansatz is given by
\bea
\delta H_3 &=& 0 \, ,  \label{dH3} \\ 
\delta F_3 &=& f_1 \, *_4 da + f_2(\tau) \, da \, \wedge \, dg^5 + 
               {\dot f}_2(\tau) \, da \, \wedge \, d\tau \, \wedge \, g^5 \, , \label{dF3} 
\\
\delta F_5 &=& (1+*) \, \delta F_3 \, \wedge \, B_2 \, , \label{dF5}
\eea
where we have used the rescaled functions as in \cite{Kuperstein:2003yt}
\bea
&& {\tilde {f}}(\tau)=- 2 P \, g_s \, f(\tau) \, , \,\,\,\,\,\,\,\,\,\,
   {\tilde {k}}(\tau)=- 2 P \, g_s \, k(\tau) \, , \,\,\,\,\,\,\,\,\,\,
   {\tilde {F}}(\tau)=  2 P \, F(\tau) \, .
\eea
$Q$ and $P$ are constants related to the number of regular and fractional
D3-branes, respectively. Indeed, $Q$ is proportional to the number of regular D3-branes, 
while $P= \frac{1}{4} \, M \, l_s^2$. We use the convention where $Q=0$
\cite{Klebanov:2000nc,Klebanov:2000hb}.
Starting from the IIB supergravity action (\ref{actionIIB}) 
it leads to the following one-dimensional effective action
\bea
S_{eff} &=& \frac{2}{\kappa^2} \, \int \left( -\frac{1}{2} \, G_{i j} \, 
{\dot {\phi}}^i \, {\dot {\phi}}^j  - V(\phi) \right) 
\,  d\tau \, .
\eea
This effective action is the same as in  \cite{Kuperstein:2003yt}. 
As before, dot stands for derivative with respect to $\tau$.
For the so-called metric in the effective action above we have the following structure
\bea
G_{i j}(\tau) \, {\dot {\phi}}^i(\tau) \, {\dot {\phi}}^j(\tau) &=& 
e^{4 p(\tau)-4 q(\tau)+4Y(\tau)} \, ( -18 \, \dot Y^2 + 45 \, \dot q^2 
+30 \, \dot p^2 + \frac{3}{2} \, \dot y^2 + \frac{3}{4} \, \dot \phi^2 + \nonumber \\
&& 3 \, \frac{\sqrt{3}}{2} \, e^{-\phi(\tau)-6 q(\tau)-4p(\tau)} \, \left(
e^{-2 y(\tau)} \, {\dot {\tilde {f}^2}} +
e^{2 y(\tau)} \, {\dot {\tilde {k}^2}} \right) + \nonumber \\
& & 3 \, \sqrt{3} \, e^{\phi(\tau)-6 q(\tau)-4p(\tau)} \,  {\dot {\tilde {F}^2}}) \, .
\label{metric}
\eea
The effective potential is given by \cite{Kuperstein:2003yt}
\bea
V(\phi) &=& e^{4 Y(\tau)} \, (\frac{1}{3} \, e^{-16 p(\tau)-4q(\tau)} 
- 2 \, e^{-6 p(\tau)-4q(\tau)} \, \cosh{y} + \frac{3}{4} \, e^{4 p(\tau)-4q(\tau)}
\, \sinh^2{y} + \nonumber \\
& & \frac{3 \sqrt{3}}{4} \, e^{\phi(\tau)-10q(\tau)+2y(\tau)} \, (2 \, P - {\tilde {F}})^2 
+ \frac{3 \sqrt{3}}{4} \, e^{\phi(\tau)-10q(\tau)-2y(\tau)} \, {\tilde {F}}^2 + \nonumber \\
& & \frac{3 \sqrt{3}}{4} \, e^{-\phi(\tau)-10q(\tau)} \, ({\tilde {k}}-{\tilde {f}})^2 +
\frac{9}{2} \, e^{-4p(\tau)-16q(\tau)} \, {\tilde {L}}^2) \, .  \label{potential} 
\eea
The potential $V(\phi)$ can be derived from the following superpotential
\bea
W &=& - 3 \,  e^{4Y(\tau)+4p(\tau)-4q(\tau)} \, \cosh{y} - 
2 \,  e^{4Y(\tau)-6p(\tau)-4q(\tau)} - 3 \sqrt{3} \,  
e^{4Y(\tau)-10q(\tau)} \, {\tilde {L}} \, , \label{superpotential}
\eea
where
\bea
V(\phi) &=& \frac{1}{8} \, G^{ij} \,  \frac{\partial W}{\partial \phi^i}
\frac{\partial W}{\partial \phi^j} \, ,
\eea
where $\phi^i=(\phi, Y, q, p, y, {\tilde {f}}, {\tilde {k}}, {\tilde {F}})$.
In addition, there is a zero-energy condition
\bea
\frac{1}{2} \, G_{i j} \, 
{\dot {\phi}}^i \, {\dot {\phi}}^j  - V(\phi) &=& 0 \, .
\eea
Therefore, the problem reduces to the finding of a non-supersymmetric deformation where 
the axionic perturbation can be treated as a perturbation of the non-supersymmetric 
deformation. This is correct since 
the ansatze for all the fields including the axionic perturbation satisfy the 
linearized equations of motion of type IIB supergravity.

Now, we show how to implement the perturbation method introduced by Borokhov and Gubser,
following the notation of Kuperstein and Sonnenschein. Consider the
effective one-dimensional Lagrangian
\bea
L &=& - \frac{1}{2} \, G_{i j} \, 
{\dot {\phi}}^i \, {\dot {\phi}}^j  - V(\phi) \, . \label{lagrangian}
\eea
The new idea introduced in \cite{Borokhov:2002fm}
consists in using the superpotential (\ref{superpotential}) to derive solutions 
satisfying the second order equations (EOM) but not the first order ones. It means that
in this case the first order equation
\bea
\frac{d \phi^i}{d \tau} &=&  \frac{1}{2} \, G^{i j} \, 
\frac{\partial W}{\partial \phi^j} \, , \label{firstorder}
\eea
is no longer valid. Now, consider a deformation of the supersymmetric
solution $\phi^i_0(\tau)$ written as
\bea
\phi^i(\tau) &=& \phi^i_0(\tau) + {\tilde \delta} \cdot {\bar {\phi^i}}(\tau) + 
{\cal {O}}({\tilde \delta}^2) \, , \label{fields}
\eea
such that $\phi^i_0(\tau)$ does satisfy Eq.(\ref{firstorder}),
where ${\tilde \delta}$ is a small positive constant that controls the non-supersymmetric 
deformation. It is conventional to introduce the functions
\bea
\zeta_i &=& G_{ij}(\phi_0) \, \left(\frac{d\bar{\phi^j}}{d\tau}-
N^j_k(\phi_0) \, \bar{\phi^k} \right) \, ,
\eea
where
\bea
N^j_k(\phi_0) &=& \frac{1}{2} \,\frac{\partial}{\partial \phi^k} \left( G^{jl}(\phi_0) \, 
\frac{\partial W}{\partial \phi^l} \right) \, .
\eea
From the second order equations derived from the Lagrangian (\ref{lagrangian}),
using the expansion (\ref{fields}) it is easy to obtain the first order differential
equation
\bea
\frac{d\zeta_i}{d\tau} &=& - N^j_i(\phi_0) \, \zeta_j  \, .
\eea
In addition, from the definition of $\zeta_i$ we have
\bea
\frac{d{\bar {\phi^i}}}{d\tau} &=& N^i_j(\phi_0) \, {\bar {\phi^j}}  +
G^{ij}(\phi_0) \, \zeta_j \, ,
\eea
while the zero-energy condition becomes
\bea
\zeta_i \, \frac{d{\bar {\phi^i}}}{d\tau} &=& 0 \, .
\eea
Therefore, the following first order differential equations are obtained
\bea
\dot \zeta_Y &=& 0 \, , \label{ode1} \\
\dot \zeta_p &=& \frac{4 \sqrt{3}}{3} \, e^{-4p_0(\tau)-6q_0(\tau)} \,
{\tilde {L}}_0  \, (\zeta_q + \zeta_Y) + e^{-10p_0(\tau)} \, 
\left( \frac{20}{9} \, \zeta_Y + \frac{8}{9} \, \zeta_q + 2 \, \zeta_p \right) , 
\label{ode2} \\
\dot \zeta_q &=&  2 \sqrt{3} \, e^{-4p_0(\tau)-6q_0(\tau)} \,
{\tilde {L}}_0  \, (\zeta_q + \zeta_Y)\, , \label{ode3} \\
\dot \zeta_y &=& - \, \left( \frac{1}{3} \, \zeta_Y + \frac{2}{15} \, \zeta_q - 
\frac{1}{5} \, \zeta_p \right) \, \sinh{y_0} + \zeta_y \, \cosh{y_0} - \nonumber \\
&& 2 \, e^{\phi_0(\tau)+2y_0(\tau)} \, ({\tilde F_0} - 2 \, P) \, \zeta_{\tilde f} - 
2 \,  e^{\phi_0(\tau)-2y_0(\tau)} \, {\tilde F_0} \, \zeta_{\tilde k} \, , \label{ode4} \\
\dot \zeta_{\tilde f+\tilde k} &=& - \frac{2 \sqrt{3}}{3} \, P \, 
e^{-4p_0(\tau)-6q_0(\tau)} \,  \, (\zeta_Y + \zeta_q) \, , \label{ode5} \\
\dot \zeta_{\tilde f-\tilde k} &=& -e^{-\phi_0(\tau)} \, \zeta_{\tilde F} +  
\frac{2 \sqrt{3}}{3} \, e^{-4p_0(\tau)-6q_0(\tau)} \, (\tilde F_0 - P) \,
(\zeta_q + \zeta_Y) \, , \label{ode6} \\
\dot \zeta_{\tilde F} &=&  -\frac{1}{\sqrt{3}} \, 
(\tilde k_0-\tilde f_0) \, e^{-4p_0(\tau)-6q_0(\tau)} \,
(\zeta_q + \zeta_Y) \nonumber \\
&& - e^{\phi_0} \, \left( \cosh(2 y_0) \, \zeta_{\tilde f-\tilde k} + \sinh(2 y_0) \,
\zeta_{\tilde f+\tilde k} \right) \, , \label{ode7} \\
\dot \zeta_\phi &=&  (2 P - {\tilde F}_0) \, e^{\phi_0(\tau)+2y_0(\tau)} \, 
\zeta_{\tilde f} + {\tilde F}_0 \, e^{\phi_0(\tau)-2y_0(\tau)} \, \zeta_{\tilde k} 
-\frac{\tilde k_0-\tilde f_0}{2} \, e^{-\phi_0(\tau)} \, \zeta_{\tilde F} \, . \label{ode8}
\eea
Notice that setting $g_s=e^{\phi}=1$, the equations above become
the ones in reference \cite{Kuperstein:2003yt}. We have used the definitions
$\zeta_{\tilde f\pm \tilde k}=\zeta_{\tilde f} \pm \zeta_{\tilde k}$. 

Now, following \cite{Kuperstein:2003yt} we show how to solve these first order
differential equations when $\zeta_Y=\zeta_p=\zeta_q=0$. Therefore, from Eq.(\ref{ode5})
$\zeta_{\tilde f + \tilde k} = X$ where $X$ is a constant of integration. From Eq.(\ref{ode6})
$\dot \zeta_{\tilde f - \tilde k} = -\zeta_{\tilde F}$. On the other hand, using  Eq.(\ref{ode4}) 
and the zero-energy condition plus the requirement of regularity in the IR, one gets 
a single solution $\zeta_{\tilde f - \tilde k} = X \, \cosh \tau$. The solutions 
for $\zeta_{\tilde f}, \,\, \zeta_{\tilde k}, \,\, \zeta_{\tilde F}, \,\, \zeta_{y}, \,\, 
\zeta_{\phi}, \,\, $ are \cite{Kuperstein:2003yt}
\bea
\zeta_{\tilde f} &=& \frac{1}{2} X (\cosh \tau +1) \, ,  \,\,\,\,\,\,
\zeta_{\tilde k} = \frac{1}{2} X (-\cosh \tau +1) \, ,  \,\,\,\,\,\,
\zeta_{\tilde F} =  -  X  \sinh\tau \, , \nonumber \\
\zeta_y &=& 2 P X  (\tau \, \cosh\tau - \sinh \tau) \, , \,\,\,\,\,\,\,\,
\dot \zeta_{\phi} = 0 \, .
\eea
Moreover, from the equation for $\phi(\tau)$ there is a unique regular solution 
for $\tau \rightarrow 0$, that corresponds to $\zeta_\phi=0$. This implies that
${\bar {\phi}}$ is a constant. In the case of ${\bar {y}}$ the
solution is
\bea
{\bar {y}}(\tau) &=& 64 \, P \, X \, 2^{2/3} \, \epsilon^{-8/3} \, \frac{1}{\sinh\tau}
\, \int_0^\tau \, \frac{x \, \coth x-1}{(\sinh(2x)-2x)^{2/3}} \, \sinh^2x \, dx \, .\label{eqy}
\eea
Using this solution, the corresponding one for ${\bar {p}}(\tau)$
is formally given by the integral
\bea
{\bar {p}}(\tau) &=& \frac{1}{5 \, \beta(\tau)}
\, \int_0^\tau \, \frac{{\bar {y}}(x)}{\sinh x} \, \beta(x) \, dx \, , \label{eqp}
\eea
where 
\bea
\beta(x) &=& e^{2 \int_{\tau_0}^\tau \, \exp[-10p_0(x)] \, dx} \, .
\eea
In the case of  ${\bar {Y}}(\tau)$ and ${\bar {q}}(\tau)$ it is convenient to solve
the first order differential equation for their difference  
${\bar {Y}}(\tau)-{\bar {q}}(\tau)$, and it gives
\bea
{\bar {Y}}(\tau)-{\bar {q}}(\tau) &=& \int_\tau^{\infty} \, 
\left( \frac{1}{5} \, \frac{{\bar {y}}(x)}{\sinh x} +
\frac{4}{3} \, e^{-10p_0(x)} \, {\bar {p}}(x) \right) \, dx \, .\label{eqY}
\eea
Still, it remains to solve the equations involving the three-form field strengths.
From Eqs.(\ref{ode5}), (\ref{ode6}) and (\ref{ode7}), and recalling the definitions
of the $\tilde f$, $\tilde k$ and $\tilde F$ in terms of $f$, $k$ and $F$, respectively,
there is the following first order ODE system
\bea
{\dot {\bar f}} + {\dot {\bar k}} + 2 \, \sinh(2y_0(\tau)) \, {\bar F}(\tau) 
- 2\, {\bar y}(\tau) \left( \dot f_0(\tau)  - 
\dot k_0(\tau) \right) &=& 2 \frac{X}{P}  h_0(\tau) \, , \nonumber \\
{\dot {\bar F}} \, \coth\tau - {\dot {\bar f}} - \frac{\coth\tau}{2} \, (\bar k - \bar f) -
e^{2y_0(\tau)} \, \bar F + 2 \, \dot f_0(\tau) \, \bar y(\tau) &=& - \frac{X}{P} h_0(\tau) 
\, , \nonumber \\
{\dot {\bar F}} \, \coth\tau + {\dot {\bar k}} - \frac{\coth\tau}{2} \, (\bar k - \bar f) -
e^{-2y_0(\tau)} \, \bar F + 2 \, \dot k_0(\tau) \, \bar y(\tau) &=& \frac{X}{P} h_0(\tau) \, ,
\label{eqsODE}
\eea
where $h_0(\tau)$ is defined in Appendix B.
In addition, there is the following first order ODE for a linear combination of $Y$ and $q$
derivatives
\bea
-2 {\dot {\bar Y}} + 5 {\dot {\bar q}} &=& \sqrt{3} \, e^{-4p_0(\tau)-6q_0(\tau)} \, \times 
\nonumber \\
&& \left( -(4 \bar p(\tau)+6 \bar q(\tau)) \, {\tilde L}_0 + (2P-{\tilde F}_0) \,
{\bar {\tilde {f}}} + {\tilde F}_0 \, {\bar {\tilde {k}}} +
({\tilde {k}}_0-{\tilde {f}}_0) \, {\bar {\tilde {F}}} \right) \, . \label{eqYq}
\eea
The formal expressions for $\bar f$, $\bar k$, $\bar F$ and $\bar q$ were
obtained in \cite{Kuperstein:2003yt} and we quote them in Appendix C.

~

Now, we can study in detail the perturbation ansatz for the 3-form and 5-form field
strengths when we consider the previously studied non-supersymmetric
deformation of the KS background. Using the ansatz (\ref{dF3}) for $\delta F_3$
we obtain
\bea
* \delta F_3 &=& \frac{1}{18 \sqrt{3}} \, e^{-q(\tau)-4p(\tau)+4Y(\tau)}
\, [f_1 \, e^{15q(\tau)-6Y(\tau)} \, da \, \wedge \, d\tau \, \wedge \, g^1 
 \, \wedge \, g^2 \, \wedge \, g^3 \, \wedge \, g^4 \, \wedge \, g^5 - \nonumber \\
&& 36 \, f_2(\tau) \,  e^{-q(\tau)-4p(\tau)-2Y(\tau)} \, (*_4 da) \, \wedge \,
d\tau \, \wedge \, dg^5 \, \wedge g^5 +  \nonumber \\
&& 81  \, {\dot f}_2(\tau) \,  e^{-q(\tau)+16p(\tau)-2Y(\tau)} \, (*_4 da) \, \wedge \,
g^1 \, \wedge \, g^2 \, \wedge \, g^3 \, \wedge \, g^4 ] \, ,
\eea
which becomes the perturbation (\ref{F30}) when the metric (\ref{metric0}) 
reduces to the Klebanov-Strassler 
background by turning off the non-supersymmetric deformation.
Similarly, using the perturbation ansatz for the five-form field strength
\bea
\delta F_5 &=& (1+*) \, \delta F_3 \, \wedge \, B_2 \, ,
\label{varF5}
\eea
together with the expression for $H_3=dB_2$
\bea
H_3 &=& -[d\tau \, \wedge \, ({\dot {\tilde {f}}}(\tau) \, g^1 \, \wedge \, g^2 +
{\dot {\tilde {k}}}(\tau) \, g^3 \, \wedge \, g^4 ) + \nonumber \\
&& \frac{1}{2} \, 
({\tilde {k}}(\tau)-{\tilde {f}}(\tau)) \, g^5 \, \wedge \, 
(g^1 \, \wedge \, g^3+g^2 \, \wedge \, g^4)] \, , 
\eea
we obtain 
\bea
\delta F_5 \, \wedge \, H_3 &=& 
f_1 \frac{(g_s \, M \, \alpha')^2}{4} \, \frac{d(f(\tau) \, k(\tau))}{d\tau} \,
(*_4 da) \, \wedge \, g^1 \, \wedge \, g^2 \, \wedge \, g^3 \, \wedge \, g^4 \, .
\eea
In order to solve the remaining equation 
$d(*\delta F_3)=\delta F_5 \, \wedge \, H_3$ we can rewrite
the perturbation of the five-form field strength (\ref{varF5}) as
\bea
\delta F_5  &=& (1+*) \, \delta {\cal {F}}_5 \, , 
\eea
where
\bea
\delta {\cal {F}}_5 &=& \delta F_3 \, \wedge \, B_2 \, ,
\eea
while for $\delta F_3$ we can write 
\bea
\delta F_3  &=& f_1 \, (*_4da) + \Delta \delta F_3 \, , 
\eea
where 
\bea
\Delta \delta F_3 = -d(f_2(\tau) \, da \, \wedge \, g^5) \, ,
\eea
is an exact form. Therefore, we get
\bea
\delta F_5 \, \wedge \, H_3  &=& (1+*) \, f_1 \, (*_4 da) \, \wedge \, 
B_2 \wedge \, H_3 \, , 
\eea
where we have used the fact that $dg^5 \, \wedge \, B_2$ and 
$d\tau \, \wedge \, g^5 \, \wedge \, H_3$ vanish.
Then, $d(*\delta F_3)=\delta F_5 \, \wedge \, H_3$ gives
the following second order ODE for the fluctuation $f_2(\tau)$ in the
non-supersymmetric deformation of the KS solution
\bea
&& - \, \frac{d}{d\tau} \left({\dot f}_2(\tau) \, e^{-2q(\tau)+12p(\tau)+2Y(\tau)} \right) +
\frac{8}{9} \, f_2(\tau) \, e^{-2q(\tau)-8p(\tau)+2Y(\tau)} = \nonumber \\
&& \frac{f_1}{12 \sqrt{3}} \, (g_s M \alpha')^2 \, 
\frac{d (f(\tau) \, k(\tau))}{d\tau} \, . \label{ODEmodified}
\eea
As expected, for ${\tilde {\delta}}=0$ this equation reduces to Eq.(\ref{ODEf20}). 
In order to solve this equation for $f_2$ 
we consider the expansion for all the fields once the non-supersymmetric deformation is 
turned on. For $f_2$ we explicitly write
\bea
f_2(\tau) &=& f_2^0(\tau) + \tilde\delta \cdot \bar f_2(\tau) + {\cal {O}}(\tilde\delta^2)
\, ,
\eea
where $f_2^0(\tau)$ is the corresponding solution (\ref{f20}) obtained by Gubser, Herzog 
and Klebanov in the supersymmetric case that we have reviewed in the previous section.
We can solve Eq.(\ref{ODEmodified}) by considering the first order perturbation expansion 
for the exponential factors  
\bea
e^{-2q(\tau)+12p(\tau)+2Y(\tau)} &=& \epsilon^{4/3} \, K^4(\tau) \, \sinh^2\tau \,  \times
 \nonumber \\
&& \left(1+ \tilde\delta \, [-2\bar q(\tau)+12\bar p(\tau)+2\bar Y(\tau)] \right) 
 + {\cal {O}}(\tilde\delta^2) \, , \\
e^{-2q(\tau)-8p(\tau)+2Y(\tau)} &=&  \frac{\epsilon^{4/3}}{K^2(\tau)}  \,  \times
\nonumber \\
&& \left(1+ \tilde\delta \, [-2\bar q(\tau)-8\bar p(\tau)+2\bar Y(\tau)] \right)  
 + {\cal {O}}(\tilde\delta^2) \, , \label{factors}
\eea
where we have replaced $q_0$, $p_0$ and $Y_0$
for their explicit functions corresponding to the KS background. Thus,
at zero order in ${\tilde \delta}$ we recover Eq.(\ref{ODEf20}). 
Therefore, we only need to solve the corresponding equation for $\bar f_2(\tau)$ as follows 
\bea
0 = -\frac{d}{d\tau} \left( K^4(\tau) \, \sinh^2\tau \, ({\dot {\bar f}}_2(\tau) +
f_2^0(\tau) \, [-2 \, \bar q(\tau)+12 \bar p(\tau) + 2 \bar Y(\tau)]) \right) + && \nonumber \\
\frac{8}{9} \, \frac{1}{K^2(\tau)} \, \left( \bar f_2(\tau) + f_2^0(\tau) \,
[-2 \, \bar q(\tau)-8 \bar p(\tau) + 2 \bar Y(\tau)]  \right) -
\frac{(g_s M \alpha')^2}{3 \epsilon^{4/3}} \, 
\frac{d(f_0 \, \bar k + k_0 \, \bar f)}{d\tau}  \, . &&
\label{ODE2}
\eea
Note that the homogeneous differential equation for $\bar f_2(\tau)$ is exactly the same
as it was for $f_2^0(\tau)$, i.e.
\bea
-\frac{d}{d\tau} [\dot {\bar f}_2(\tau) \, K^4(\tau) \, \sinh^2\tau ] + 
\frac{8 \, \bar f_2(\tau)}{9 \, K^2(\tau)} &=& 
0 \, .
\eea
As in the supersymmetric case this is solved by the functions
\bea
\bar f_2^{(1)}(\tau) &=& [\sinh(2 \tau) - 2 \tau]^{1/3} \, , \\
\bar f_2^{(2)}(\tau) &=& [\sinh(2 \tau) - 2 \tau]^{-2/3} \, .
\eea
To solve the inhomogeneous equation we use the Wronskian 
\bea
W(\bar f_2^{(1)}(\tau), \bar f_2^{(2)}(\tau)) &=& -
\frac{4 \sinh^2\tau}{[\sinh(2\tau)-2\tau]^{4/3}} \, .
\eea
Then, we recast Eq.(\ref{ODE2})
\bea
&& \frac{d^2{\bar f}_2(\tau)}{d\tau^2} + \left(
 \frac{1}{K^4(\tau) \, \sinh^2\tau} \, 
\frac{d}{d\tau} (K^4(\tau) \, \sinh^2\tau ) \right)
\, \frac{d{\bar f}_2(\tau)}{d\tau} - 
\frac{8}{9 \, K^6(\tau) \, \sinh^2\tau} \, 
\bar f_2(\tau) \nonumber \\
&& = \bar {j}(\tau) \, ,
\eea
such that now the source becomes
\bea
{\bar {j}}(\tau) &=& [-\eta \, \frac{d}{d\tau}(f_0(\tau) \, 
\bar k(\tau) + k_0(\tau) \, \bar f(\tau))
+ \frac{8}{9 \, K^2(\tau)} \, f^0_2(\tau) \, 
[-2 \bar q(\tau) - 8 \bar p(\tau) + 2 \bar Y(\tau)] - \nonumber \\
&& \frac{d}{d\tau} \left( K^4(\tau) \, \sinh^2\tau \,\, f_2^0(\tau) \, 
[-2 \bar q(\tau) + 12 \bar p(\tau) + 2 \bar Y(\tau)]
\right) ] \, \frac{1}{K^4 \, \sinh^2\tau}  \, ,
\eea
where $\eta = \frac{(g_s M \alpha')^2}{3 \epsilon^{4/3}}$. The general solution is
\bea
\bar f_2(\tau) &=& d_1 \cdot \bar f_2^{(1)}(\tau) + d_2 \cdot \bar f_2^{(2)}(\tau)
 - \bar f_2^{(1)}(\tau) \, \int_0^\tau \, \bar f_2^{(2)}(x) \, 
\frac{{\bar {j}}(x)}{W(\bar f_2^{(1)}(x), \bar f_2^{(2)}(x))} \, dx + \nonumber \\
&&  \bar f_2^{(2)}(\tau) \,
\int_0^\tau \, \bar f_2^{(1)}(x) \, 
\frac{{\bar {j}}(x)}{W(\bar f_2^{(1)}(x), \bar f_2^{(2)}(x))} \, dx \, , \label{barf2}
\eea
where $d_1$ and $d_2$ are constants determined by the boundary conditions.
In the present case in order to ensure the finiteness of the solution at 
$\tau \rightarrow 0$ we require $d_2=0$, while for $\tau \rightarrow \infty$
it implies that 
\bea
d_1 &=& \frac{891}{32} \, \frac{(g_s M \alpha')^3}{2^{2/3}} \, \frac{X}{\epsilon^4}
\, .
\eea
We are interested in analyzing the asymptotic behaviour of $f_2(\tau)$,
both at zero and infinity, in order to determine whether the supergravity mode
associated with the perturbation ansatz given by Eqs.(\ref{dH3}), (\ref{dF3}) and (\ref{dF5}) is normalizable.

~

{\it Asymptotic behaviour of $f_2(\tau)$ for $\tau \rightarrow 0$.}

~

Let us consider the asymptotic behaviour for the previously studied fields
when $\tau \rightarrow 0$. They lead to the following expressions
\bea
h_0(\tau) & \approx & 2^{2/3} \, \frac{(4P)^2}{\epsilon^{8/3}} \, (a_0-a_1 \, \tau^2)+
{\cal {O}}(\tau^3) \, , \\
\bar y(\tau) &\approx& \frac{3^{2/3}}{27} \, \mu \, \tau^2 + {\cal {O}}(\tau^4) \, ,\\
\bar p(\tau) &\approx& \frac{3^{2/3}}{675} \, \mu \, \tau^2 + {\cal {O}}(\tau^4) \, ,\\
\bar q(\tau) &\approx& \frac{3^{2/3} \cdot 11}{4050} \, \mu \, \tau^2 + {\cal {O}}(\tau^4) \, ,\\
\bar Y(\tau) &\approx& C_Y^0 - \frac{3^{2/3}}{405} \, \mu \, \tau^2 + {\cal {O}}(\tau^4) \, , \\
\bar F(\tau) &\approx& \gamma \, \tau^2 +  {\cal {O}}(\tau^4) \, ,\\
\bar f(\tau) &\approx& \frac{1}{2} \, \gamma \, \tau^3 +  {\cal {O}}(\tau^5) \, ,\\
\bar k(\tau) &\approx& -2 \, \gamma \, \tau +  {\cal {O}}(\tau^3) \, ,
\eea
where $a_0$, $a_1$, $C_Y^0$ are constants, while
\bea
\mu &=& 96 \cdot 2^{1/3} \, g_s P X \epsilon^{-8/3} \, ,
\,\,\,\,\,\,\,\,\,\,\,\,\,
\gamma = - \frac{2^{1/3}}{18} \, \mu \, a_0 \, .
\eea
As we have already seen from the supersymmetric case there is
an integral expression for $f_2^0(\tau)$ which is well behaved for $\tau \rightarrow 0$.
We explicitly obtain
\bea
f_2^0(\tau) &=& \frac{a_0 (g_s M \alpha')^2}{6 \cdot 3^{1/3} \epsilon^{4/3}} \, \tau
- \frac{a_0 (g_s M \alpha')^2}{45 \cdot 3^{1/3} \epsilon^{4/3}} \, \tau^3
+ {\cal {O}}(\tau^5)\, ,
\eea
while for $\bar f_2(\tau)$ Eq.(\ref{barf2}) reduces to
\bea
\bar f_2(\tau) &=& \frac{(g_s M \alpha')^3 \, X}{\epsilon^{4}} \, \left( \frac{297 \cdot 3^{2/3}}{32} \, \tau
+ \left( \frac{3^{2/3} \, \cdot \, 99}{160} -
\frac{2^{1/3} \, \cdot \, 4 \, a_0}{3^{2/3} \, \cdot \,  75}
 \right) \, \tau^3 \right) + {\cal {O}}(\tau^4)\, ,
\eea
which in the low energy expansion for $f_2(\tau)$ falls with the same power of $\tau$
as the leading term of $f_2^0(\tau)$.

~

{\it Asymptotic behaviour of $f_2(\tau)$ for $\tau \rightarrow \infty$.}

~

Now, we consider the opposite asymptotic limit of the fields, i.e. when 
$\tau \rightarrow \infty$. The corresponding expressions are
\bea
h_0(\tau) & \approx & 2^{1/3} \cdot \frac{3}{4} \, \alpha \, \tau e^{-4 \tau/3} + 
\cdot \cdot \cdot \, , \\
\bar y(\tau) &\approx& \mu \, \left(\tau-\frac{5}{2} \right) \, e^{-\tau/3} + V \, e^{-\tau}
+  \cdot \cdot \cdot \, , \\
\bar p(\tau) &\approx& \frac{3}{5} \, \mu \, \left(\tau-4 \right) \, e^{-4 \tau/3} 
+  \cdot \cdot \cdot \, , \\
\bar q(\tau) &\approx& -\frac{2}{5} \, \mu \, \tau \, e^{-4 \tau/3} +  \cdot \cdot \cdot \, , \\
\bar Y(\tau) &\approx&  \mu \, \left( \frac{1}{2} \tau - \frac{99}{40} \right) \,
  e^{-4 \tau/3} +  \cdot \cdot \cdot \, , \\
\bar F(\tau) &\approx& 3 \, \mu \, \left( \frac{1}{4} \tau - 1 \right) \, e^{-\tau/3}
+  \left( \frac{3 V}{2}+V'\right) \, e^{-\tau} +  {\cal {O}}(e^{-4 \tau/3}) \, ,\\
\bar f(\tau) &\approx& -\frac{27}{16} \, \mu \,  e^{-\tau/3} + 
\left( \frac{V}{2}+V'\right) \, e^{-\tau}  +  {\cal {O}}(e^{-4 \tau/3}) \, ,\\
\bar k(\tau) &\approx& \frac{27}{16} \, \mu \,  e^{-\tau/3} - 
\left( \frac{V}{2}+V'\right) \, e^{-\tau}  +  {\cal {O}}(e^{-4 \tau/3}) \, ,
\eea
where $\alpha=4 (g_s M \alpha')^2 \, \epsilon^{-8/3}$, while $V$ and $V'$ are constants.

In addition, we have
\bea
f_2^0(\tau) &=& \frac{3 (g_s M \alpha')^2}{4 \cdot 2^{1/3}  \, \epsilon^{4/3}} \, \tau \,
e^{-2\tau/3} + {\cal {O}}(\tau \, e^{-8\tau/3}) \, .
\eea
In the limit $\tau \rightarrow \infty$  Eq.(\ref{barf2}) reduces to
\bea
\bar f_2(\tau) &=& \frac{891 \, (g_s M \alpha')^3 \, X}{\epsilon^4} 
 \, e^{-4\tau/3}  + {\cal {O}}(\tau^2 \, e^{-2\tau}) \, .
\eea

Therefore, we have explicitly shown that the supergravity mode associated with
the perturbation ansatz is well-behaved both in the IR and UV limits.

~

{\it Asymptotic behaviour of $\delta  F_3$}

~

Now, we study the normalizability of $\delta  F_3$. As in the previous 
section we must integrate
\bea
\int \, \sqrt{|g|} \, |\delta F_3|^2 \, d^{10}x &=& 
\int \, \delta F_3 \, \wedge \, *\delta F_3 \, ,
\eea
using the deformed solutions that we have obtained in this section. The above integral
can be written as the sum of the following ones
\bea
I_1^0 &=& - f_1^2 \, \frac{\epsilon^4}{96} \, 
\int_0^\infty \, h_0^2(\tau) \, \sinh^2\tau \, d\tau  \\
\bar I_1 &=& - f_1^2 \, \frac{\epsilon^4}{96} \, 
\int_0^\infty \, h_0^2(\tau) \, \sinh^2\tau \, 
[14 \bar q(\tau) - 4 \bar p(\tau) - 2 \bar Y(\tau)] \, d\tau \, , \\
I_2^0 &=&  \frac{\epsilon^{4/3}}{3} \, \int_0^\infty \, \frac{(f_2^0(\tau))^2}{K^2(\tau)} 
\, d\tau \, , \\
\bar I_2 &=& \frac{\epsilon^{4/3}}{3} \, \int_0^\infty \, \frac{1}{K^2(\tau)} \,
\left((f_2^0(\tau))^2 \, [-2 \bar q(\tau) -8 \bar p(\tau) + 2 \bar Y(\tau)] + 
2 f_2^0(\tau) \, \bar f_2(\tau) \right)  \, d\tau\, , \\
I_3^0 &=& \frac{3 \, \epsilon^{4/3}}{8} \, \int_0^\infty \, ({\dot f}_2^0(\tau))^2  \, K^4(\tau) 
\, \sinh^2\tau \, d\tau \, , \\
\bar I_3 &=& \frac{3 \, \epsilon^{4/3}}{8} \, \int_0^\infty \, K^4(\tau) 
\, \sinh^2\tau \,  \times  \nonumber \\
&& \left( ({\dot f}_2^0(\tau))^2 \, [-2 \bar q(\tau) + 12 \bar p(\tau) 
+ 2 \bar Y(\tau)] + 2 {\dot f}_2^0(\tau) \, {\dot {\bar f}}_2(\tau)  \right)\, d\tau \, .
\eea
We have written the integrals $I_i^0$ and $\bar I_i$ above according to the expansion  
$I_i=I^0_i + \tilde \delta \cdot \bar I_i + {\cal {O}}(\tilde \delta^2)$, for 
$i=1$, $2$ and $3$. These integrals are well-behaved in the IR limit. All the integrals
$I^0_i$ fall as $e^{-2\tau/3}$ in the UV limit (when $\tau$ becomes large), while
$\bar I_i$'s fall even faster, as $e^{-2\tau}$, in the UV limit.

This shows that we have found a normalizable zero-mode for the
non-supersymmetric deformation of the warped deformed conifold. 
This is associated with a massless pseudo-scalar glueball
in the non-su\-per\-symme\-tric gauge theory. We will
return to this result in the next section.

It is a trivial check to show how 
to get the supersymmetric perturbation obtained by Gubser, Herzog and Klebanov
reviewed in the previous section by just taking the non-su\-per\-symme\-tric
deformation parameter $\tilde \delta=0$. This, of course, sets the situation back to
the supersymmetric Klebanov-Strassler background since $\tilde \delta=0$ means
that the solutions to the second order equations of motion satisfy the first
order ones. In such a case $f_2(\tau)$ becomes $f_2^0(\tau)$, being  
well-behaved both at zero and infinity, and the perturbation to the three-form 
field strength is also normalizable.

\section{Zero modes, glueballs and symmetry breaking in
supersymmetric and non-su\-per\-sy\-mme\-tric dual gauge theories}

In this section we study properties of the dual field theory 
associated with the axionic non-supersymmetric deformation that we have discussed
in the previous section, and perform a parallel analysis to the case discussed
by Gubser, Herzog and Klebanov. 

Since our solution is based on the one reported in \cite{Kuperstein:2003yt}, it is a regular,
non-su\-per\-symme\-tric first order deformation of the KS solution which has its
same isometries. Thus, in our solution the deformation corresponds to the
inclusion of a mass term of the gaugino bilinears in the dual field theory.
The KS background has a well-known dual description in terms a four dimensional
${\cal {N}}=1$ supersymmetric Yang Mills theory with the $SU(N+M) \times SU(N)$ gauge group, 
where on the supergravity side $N$ and $M$ are the numbers of regular and fractional
D3-branes, respectively. This gauge theory is coupled to four bifundamental chiral
multiplets $A_i$ and $B_j$, with $i$ and $j=1$, $2$ transforming under $SU(N+M) \times SU(N)$
gauge group.
In addition, each set of fields $A_i$ and $B_j$ transforms as a doublet under the action
of one of the two $SU(2)$'s in the $SU(2) \times SU(2)$ global symmetry group. 
The theory is believed to undergo a cascade of Seiberg dualities
towards the IR. As pointed out in \cite{Gubser:2004qj}, the cascade stops
leading to the $SU(2M) \times SU(M)$ gauge group. The KS background has a deformation
parameter $a_0 \sim \epsilon^{-8/3}$, such that it is related to a four dimensional mass
scale $m \sim \epsilon^{2/3}$. For a non-vanishing $\epsilon$ the $U(1)_R$ symmetry of
the conifold is broken down to ${\bf Z}_2$, which is preserved by the gaugino bilinear
$Tr \lambda \, \lambda$. In this case the potential that appears in the effective action
has a critical point corresponding to the superconformal theory dual to the 
$AdS_5 \times T^{1,1}$ background, where there are no fractional D3-branes.
Now, if one expands the potential around the critical point and uses the usual 
mass/dimension formula
of the AdS/CFT correspondence, one gets the dimension of the fields that can be
identified with certain gauge theory operators \cite{Ceresole:1999zs,Bigazzi:2000uu}. 
In particular, two of these
operators are $Tr(W^2_{(1)})-W^2_{(2)})$ (which on the
supergravity side corresponds to the mode $y(\tau)$), and $Tr(W^2_{(1)})+W^2_{(2)})$,
(associated with $\zeta_2(\tau) \sim -F(\tau) + (k(\tau)-f(\tau))/2$), having both
dimension $\Delta=3$. The non-supersymmetric deformation studied by Kuperstein and Sonnenschein
was obtained by introducing mass terms of the gaugino bilinears associated
with $\zeta_2$ and $y$. The most general gaugino bilinear has the form of
$\mu_+ {\cal {O}}_+ + \mu_- {\cal {O}}_- + c.c.$, where 
${\cal {O}}_\pm \sim Tr(W^2_{(1)}) \pm W^2_{(2)})$, and $W_{(i)}$ with $i=1$, $2$
labels the $SU(N+M)$ and $SU(N)$ gauge groups, respectively. The deformation 
\cite{Kuperstein:2003yt} has only one real parameter $\mu$. All the fields behave
regularly in the IR and in the UV limits, however, as mentioned before there are
two non-normalizable modes which are precisely $y(\tau)$ and $\zeta_2(\tau)$.
The mass term induces the so-called soft supersymmetry 
breaking. One should notice that, as was pointed out in \cite{Kuperstein:2003yt}, 
when in the deformed solution (\ref{Ffk}) the constants are chosen as 
$C_1=-1/2$ while $C_2=C_3=0$, a $(0, 3)$ form is obtained (see Appendix C). 
This breaks the supersymmetry 
and the solution diverges at $\tau \rightarrow \infty$. On the other hand, the
difference between the vacuum energy of the deformed non-supersymmetric theory
and the corresponding one of the supersymmetric Klebanov-Strassler solution is finite.

In the previous section we have obtained a normalizable zero mode in a 
non-su\-per\-symme\-tric deformation of the Klebanov-Strassler background. 
The presence of such a zero mode is related to the properties of the 
non-supersymmetric deformation inherited from the KS solution, and it is
related to a spontaneously broken symmetry. Particularly, as in the KS case
we would like to argue that in the dual field theory this massless mode
is identified with a massless pseudo-scalar glueball generated by the
spontaneously broken $U(1)_B$ baryon number symmetry\footnote{We should
emphasize that apart from the massless glueballs found in \cite{Gubser:2004qj}, in
previous calculations \cite{Krasnitz:2000ir,Caceres:2000qe,Amador:2004pz}
of glueball spectra no massless modes were found.}. An important point here
is the fact that in the framework of the AdS/CFT duality global symmetries in the  
field theory become gauge symmetries in the dual supergravity theory. Now, in order
to identify the gauge field on the field theory description with the massless pseudo-scalar
fluctuation in the type IIB supergravity description the argument given in \cite{Gubser:2004qj} 
does straightforwardly apply. Firstly, let us remember that in the case
of the dual superconformal fixed point of the conifold background solution 
\cite{Klebanov:1998hh} the global symmetries 
are $SU(2) \times SU(2) \times U(1)_R \times U(1)_B$, which are all preserved.
In particular, in \cite{Klebanov:1999tb,Ceresole:1999zs} the gauge field $A$ that is dual to the
baryon number current $J^\mu$ was identified as $\delta C_4 \sim \omega_3 \wedge A$.

On the other hand, when we consider the Klebanov-Strassler solution the
$SU(2) \times SU(2)$ global symmetry is preserved, while the global $U(1)_R$ 
symmetry group breaks down to
${\bf Z}_{2M}$ in the UV, due to the chiral anomaly. This ${\bf Z}_{2M}$ symmetry 
spontaneously breaks down to ${\bf Z}_2$. There is no Goldstone boson associated with this
symmetry breaking. 

The axionic perturbation is a massless speudo-scalar glueball interpreted as the 
Goldstone boson of spontaneously $U(1)_B$ baryon number symmetry.
From the structure of the five-form field strength perturbation in Eq.(\ref{dF50}) one can
read off the relation between the supergravity zero-mode and the glueball, since
in the UV limit there is a component in $\delta F_5$ which behaves as
$\omega_3 \wedge da \wedge d\tau$, leading to the identification of gauge field $A$
with $da$. In the non-supersymmetric deformation the same comments apply, and in both cases
the effective four-dimensional Lagrangian is
\bea
\frac{1}{f_a} \, \int d^4 x \, J^\mu(x) \, \partial_\mu a(x) &=&
- \frac{1}{f_a} \, \int d^4 x \, a(x) \, (\partial_\mu \, J^\mu(x)) \, .
\eea
Particularly, in the non-supersymmetric deformation we can easily identify the
supergravity mode associated with the $U(1)_B$, just looking at the UV limit of the  
Eq.(\ref{dF5}), where we also find a component with structure 
like $\omega_3 \wedge da \wedge d\tau$,
that can be used to identify the gauge field $A$ with $da$.

The theory is on the baryonic branch since the $SU(2) \times SU(2)$ global symmetry 
is preserved \cite{Klebanov:2000hb,Aharony:2000pp,Gubser:2004qj}, and the only operators 
which preserve such a symmetry are the baryonic operators, so that they do have
non-vanishing VEVs.
In the supersymmetric field theory the series of Seiberg dualities stops at
the theory with gauge group $SU(2M) \times SU(M)$, which is coupled to
bifundamental fields $A_i$ and $B_j$, $i$, $j=1$, $2$. The $SU(2M) \times SU(M)$ 
gauge invariant baryonic operators in the field theory are 
\bea
{\cal B} &\approx& \epsilon_{\alpha_1 \alpha_2 \cdot \cdot \cdot \alpha_{2M}} \,
(A_1)_1^{\alpha_1} (A_1)_2^{\alpha_2} \cdot \cdot \cdot 
(A_1)_M^{\alpha_M} (A_2)_1^{\alpha_{M+1}} (A_2)_2^{\alpha_{M+2}} 
 \cdot \cdot \cdot (A_2)_M^{\alpha_{2M}} \, ,  \\
\bar {\cal B} &\approx& \epsilon^{\alpha_1 \alpha_2 \cdot \cdot \cdot \alpha_{2M}} \,
(B_1)^1_{\alpha_1} (B_1)^2_{\alpha_2} \cdot \cdot \cdot 
(B_1)^M_{\alpha_M} (B_2)^1_{\alpha_{M+1}} (B_2)^2_{\alpha_{M+2}} 
 \cdot \cdot \cdot (B_2)^M_{\alpha_{2M}} \, .  
\eea
This theory has a superpotential, such that the baryonic branch is one of the supersymmetric 
background solutions where $SU(2) \times SU(2)$ global symmetry is preserved.
It leads to identify the dual of this background with the baryonic branch of the 
cascading theory. On the other hand, the expectation values of the baryonic
operators spontaneously break the $U(1)_B$ baryon number symmetry. The vacuum corresponding
to the deformed conifold has VEVs for the baryonic operators such as
$|{\cal B}|=|\bar {\cal B}|=\Lambda^{2M}_{2M}$. Essentially, the baryonic branch
can be parametrized by $\xi$, such that ${\cal B}=i \xi \Lambda^{2M}_{2M}$ and
$\bar {\cal B}=i \xi^{-1} \Lambda^{2M}_{2M}$. Therefore, the pseudo-scalar Goldstone boson
must correspond to change  $\xi$ by a phase. This is because the baryon number symmetry
transforms the bifundamentals as $A_i \rightarrow e^{i \alpha} A_i$ and 
$B_j \rightarrow e^{-i \alpha} B_j$. Then, $f_a \partial_\mu a(x)$ is created through
the action of the axial baryon number current $J_\mu$ on the vacuum. Precisely, 
$f_a \partial_\mu a(x)$ is the gradient of the pseudo-scalar Goldstone boson.

In the supersymmetric gauge theory it is expected to find a massless scalar field,
the saxion, which was explicitly identified with its dual supergravity
fluctuation mode in \cite{Gubser:2004qj}. In addition, as remarked in 
\cite{Gubser:2004qj} the original ${\bf Z}_2$ symmetry related to the interchange 
of the two $S^2$'s 
in the base of the singular conifold is preserved by the warped deformed conifold metric,
as well as the $F_5$, while $F_3$ and $H_3$ change sign. In the dual gauge theory
it leads to the interchange of the two doublets of bifundamentals. The axion, both
in the supersymmetric and non-supersymmetric cases, breaks this ${\bf Z}_2$ symmetry. We can
see this from the perturbation ansatze (\ref{dF50}) and (\ref{dF5}), respectively.

It would be interesting to extend these studies to non-supersymmetric 
deformations of different backgrounds. For instance, the solution obtained by Pando Zayas and
Tseytlin \cite{PandoZayas:2000sq} has the same asymptotic behaviour as the KS one. 
However, this solution
deals with a resolved conifold instead of the deformed one. In \cite{Cvetic:2000db} it was 
pointed out that this solution is not supersymmetric. This solution is singular
and has a repulson. It would be possible to resolve such a singularity through 
an enhancon mechanism, replacing it by some distribution of branes consistent with
this background. On the other hand, as it was proposed in \cite{Aharony:2000pp} it might 
be possible that after resolving the singularity, the solution becomes supersymmetric.
In any case, an interesting point is that the field theory is conjectured to be
on a mesonic branch of the moduli space, 
where the mesonic operators would break the $SU(2) \times SU(2)$ global symmetry.
Therefore, it would certainly be of interest to identify the supergravity modes
associated with the Goldstone bosons related to the spontaneously broken  
$SU(2) \times SU(2)$ global symmetry. The fact that in this case the 2-cycle 
does not collapse may suggest that the $U(1)_B$ baryon number symmetry would remain unbroken.
These are matters of conjecture and deserve to be carefully studied.

~

\section{Discussion and conclusions}

We have obtained an explicit solution for an axionic perturbation of  
the non-su\-per\-symme\-tric deformation of the warped deformed conifold of 
Kuperstein-Sonnenschein. The corresponding supergravity fluctuation is normalizable, and
we have performed an analysis analogous to the one developed by Gubser,
Herzog and Klebanov in the supersymmetric warped deformed conifold.
This is the Goldstone boson associated with the spontaneous breaking of the $U(1)_B$ 
baryon number symmetry in the dual gauge theory of the non-supersymmetric warped deformed conifold 
background. Moreover, the gauge theory is on the baryonic branch of the moduli space
since the $SU(2) \times SU(2)$ symmetry of the dual supergravity
background is preserved, and therefore the only operators with non-vanishing
vacuum expectation values are the baryonic ones.
Still, there are some points that deserve further investigation.
It was argued that the stability of the non-supersymmetric
solution requires a mass gap in the original dual supersymmetric gauge theory \cite{Aharony:2002vp}. 
On the other hand,
Gubser, Herzog and Klebanov have shown that in the Klebanov-Strassler solution there 
is no mass gap. Therefore,  a very important question arises, i.e. whether
the non-supersymmetric deformation of Kuperstein and Sonnenschein induces tachyonic
modes, as well as if the saxion becomes a tachyonic mode itself.   
In addition, it would be useful to discuss the stability of the D1-branes 
at the bottom of the non-supersymmetric background. It would be expected that they  
were stable for small values of the non-supersymmetric deformation parameter.

On the other hand, although in the Maldacena-N\'u\~nez solution 
the integral $\int \delta F_3 \wedge *\delta F_3$ has a divergence
that may be related to the linear dilaton background associated with
the D5-brane in the UV, it would be interesting both, 
to find a normalizability criterion
for the supergravity mode associated with the axion field perturbation, as well as
to extend the analysis in the lines proposed here for non-supersymmetric deformations
of this solution \cite{Aharony:2002vp,Evans:2002mc,Gubser:2001eg,Bertoldi:2004rn}. 
As suggested in \cite{Gubser:2004qj}, perhaps it may be possible to understand
D-strings in the IR of the solution \cite{Maldacena:2000yy} as axionic strings. 

It would also be interesting to obtain explicitly the supergravity fluctuation modes
associated with the would-be mesonic branch of the moduli space 
related to the dual gauge theory of the Pando Zayas-Tseytlin background.


~

~


\centerline{\large{\bf Acknowledgments}}

~

We are grateful to Vicente Di Clemente for useful discussions and Sebastian Franco
for a critical reading of the manuscript.
This work has been supported by the PPARC
Grant PPA/G/O/2000/00469 and the Fundaci\'on Antorchas.

\newpage

\subsection*{Appendix A: Type IIB supergravity in ten dimensions}

In order to define our notation we briefly review the action and
equations of motion corresponding to ten dimensional type IIB
supergravity \cite{Schwarz:qr}. The field content of this theory
is given by the metric, a 4-form potential $C_4$, a scalar
$\phi$, an axion $\chi$, a R-R 2-form potential $C_2$,
a NS-NS 2-form potential $B_2$, two gravitinos with the
same chirality $\Psi^i_M$, and two dilatinos $\lambda_i$
($i=1,2$). Since there is not a simple covariant Lagrangian for
type IIB supergravity under the condition $F_5=*F_5$, one can
write a Lagrangian without constraining the five-form field
strength and, after derivation of the equations of motion, one can
impose that condition \cite{Bergshoeff:1995as}. We use the
notation given in \cite{Kuperstein:2003yt} (see also \cite{Tran:2001gw}). 
We consider the bosonic type IIB supergravity action written in the Einstein
frame
\bea
S_{IIB} &=&  \frac{1}{2 \, \kappa^2} \int d^{10}x \,
\sqrt{-g} \, R_E - \frac{1}{4 \, \kappa^2} \int d^{10}x \,
[ d\phi \wedge *d\phi + e^{2 \phi} \, d\chi \wedge *d\chi + \nonumber \\
& & g_s \, e^{- \phi} \, H_3 \wedge *H_3 + 
g_s \, e^{\phi} \, F_3 \wedge *F_3 + 
\frac{g_s^2}{2} \, F_5 \wedge *F_5 +
g_s^2 \, C_4 \wedge H_3 \wedge F_3 ] \, ,
\label{actionIIB}
\eea
with usual definitions for the fields.
The equations of motion derived from the action (\ref{actionIIB}) are
\bea
R_{MN}&=&\frac{1}{2} \, \partial_M \phi \, \partial_N \phi +
\frac{1}{2} \, e^{2 \phi} \, \partial_M \chi \, \partial_N \chi +
+ \frac{g_s^2}{96} \, F_{MPQRS} \, F_N^{\,\,\,PQRS} \nonumber
\\
&& + \frac{g_s}{4} \, e^\phi \, (F_{MRS} \, F_N^{\,\,\,RS} -
\frac{1}{12} \, F_{RST} \, F^{RST} \, g_{MN}) \nonumber \\
&& + \frac{g_s}{4} \, e^{-\phi} \, (H_{MRS} \, H_N^{\,\,\,RS} -
\frac{1}{12} \, H_{RST} \, H^{RST} \, g_{MN}) \, ,
\label{eom-einstein-IIb} \\
d * d\phi &=&   e^{2 \phi} \, d\chi \, \wedge \, *d\chi +
\frac{g_s \, e^{\phi}}{2} \, F_3 \, \wedge \, *F_3 -
\frac{g_s e^{-\phi}}{2} \, H_3 \, \wedge \, *H_3  \, ,
\label{eom-dilaton-IIb}
\\
d(e^{2 \phi} \, * d\chi) &=& -g_s \, e^\phi \, H_3 \, \wedge \, *F_3
\, , \label{eom-axion-IIb} \\
d(e^{\phi} \, * F_3) &=& g_s \, F_5 \, \wedge \, H_3 \, , \label{eom-f3}
\\
d*(e^{- \phi} \, H_3 - e^\phi \, \chi \, F_3) &=& -  g_s \, F_5 \,
\wedge \, F_3  \, .  \label{eom-h3} 
\eea
In addition, there are the following Bianchi identities
\bea
dF_3 &=& -d\chi \, \wedge H_3 \, , \label{bianchi0} \\
dF_5 &=& H_3 \, \wedge F_3 \, , \label{bianchi1} \\
dH_3 &=& 0 \, . \label{bianchi2} 
\eea
In this paper the ten dimensional axion $\chi$ of type IIB supergravity is set to zero, while
we call axionic perturbation to a perturbation included in the fluctuation spectrum 
of type IIB supergravity defined as a particular combination of perturbations of the three-form 
and five-form field strengths (see Sections 2 and 3).  

~

\subsection*{Appendix B: A collection of formulas of the 
Klebanov-Strassler solution}

We write some explicit formulas of \cite{Klebanov:2000hb} which are relevant 
for the calculations we have presented. We have used the notation given in
\cite{Gubser:2004qj}, \cite{Herzog:2001xk} and \cite{Herzog:2002ih}.
The 10d metric is
\bea
ds^2_{10} &=& h_0(\tau)^{-1/2} \, (-dt^2 + d{\vec {x}}_3^2) +
              h_0(\tau)^{1/2} \, ds^2_{6} \, , \label{metricwarped}
\eea
where the warped deformed conifold metric is given by
\bea
ds^2_{6} &=& \frac{\epsilon^{4/3}}{2} \, K(\tau) \, [ \frac{1}{3 K(\tau)^3}
             \, (d\tau^2 + (g^5)^2) + \nonumber \\
         & & \cosh^2 \left( \frac{\tau}{2} \right) \, ((g^3)^2+(g^4)^2) +
             \sinh^2 \left( \frac{\tau}{2} \right) \, ((g^1)^2+(g^2)^2)] \, ,
\eea
where
\bea
K(\tau) &=& \frac{(\sinh(2 \tau) - 2 \tau)^{1/3}}{2^{1/3} \, \sinh\tau} \, , \\
        & & \nonumber \\
h_0(\tau) &=& (g_s M  \alpha')^2 \, 2^{2/3} \, \epsilon^{-8/3} \, I(\tau) \, , \\
        & & \nonumber \\
I(\tau) &=& \int_\tau^\infty \, dx \, \frac{x \, \coth x - 1}{\sinh^2x} \, 
            (\sinh(2 x)- 2 x)^{1/3}  \, .
\eea
The one-forms are
\bea
g^1 &=& \frac{e^1-e^3}{\sqrt{2}} \, , \,\,\,\,\,\,\,\,\, 
g^2 = \frac{e^2-e^4}{\sqrt{2}} \, , \nonumber \\
g^3 &=& \frac{e^1+e^3}{\sqrt{2}} \, , \,\,\,\,\,\,\,\,\, 
g^4 = \frac{e^2+e^4}{\sqrt{2}} \, , \nonumber \\
e^5 &=& g^5 \, ,
\eea
where
\bea
e^1 & \equiv & -\sin\theta_1 \, d\phi_1 \, , \nonumber \\
e^2 & \equiv & d\theta_1 \, , \nonumber \\
e^3 & \equiv & \cos\psi \, \sin\theta_2 \, d\phi_2 - \sin\psi \, d\theta_2 \, , \nonumber \\
e^4 & \equiv & \sin\psi \, \sin\theta_2 \, d\phi_2 + \cos\psi \, d\theta_2 \, , \nonumber \\
e^5 & \equiv & d\psi + \cos\theta_1 \, d\phi_1 + \cos\theta_2 \, d\phi_2 \, .
\eea
The NS-NS two form field is
\bea
B_2 &=& \frac{g_s M \alpha'}{2} \, [f(\tau) \, g^1 \, \wedge \, g^2 +
                                          k(\tau) \, g^3 \, \wedge \, g^4] \, ,
\eea
while its corresponding three form field strength is given by 
\bea
H_3 &=& dB_2 = \frac{g_s M \alpha'}{2} \, 
[d\tau \, \wedge \, ({\dot f}(\tau) \, g^1 \, \wedge \, g^2 +
{\dot k}(\tau) \, g^3 \, \wedge \, g^4) + \nonumber \\
& & \frac{1}{2} \, (k(\tau) - f(\tau)) \, g^5 \, \wedge \, (g^1 \, 
\wedge \, g^3 + g^2 \, \wedge \, g^4)]        \, .
\eea
The R-R three form field strength is 
\bea
F_3 &=& \frac{M \alpha'}{2} \, 
[(1-F(\tau)) \, g^5 \, \wedge \, g^3 \, \wedge \, g^4 + 
F(\tau) \, g^5 \, \wedge \, g^1 \, \wedge \, g^2 + \nonumber \\
& & {\dot F}(\tau) \, d\tau \, \wedge \, ( g^1 \, \wedge \, g^3 + g^2 \, \wedge \, g^4 )] \, .
\eea
In addition 
\bea
F(\tau) &=& \frac{\sinh\tau-\tau}{2 \, \sinh\tau} \, , \\ 
        & &  \nonumber \\
f(\tau) &=& \frac{\tau \, \coth\tau-1}{2 \, \sinh\tau} \, (\cosh\tau-1) \, , \\ 
        & &  \nonumber \\
k(\tau) &=& \frac{\tau \, \coth\tau-1}{2 \, \sinh\tau} \, (\cosh\tau+1) \, , 
\eea
where they correspond to the first term in each of the equations in (\ref{Ffk}).
Also, we introduce the definition two-form
\bea
\omega_2 &=& \frac{1}{2} \, (g^1 \, \wedge \, g^2 + g^3 \, \wedge \, g^4 ) \, , 
      \,\,\,\,\,\,\,          \,\,\,\,\,\,\, \omega_3 = g^5 \, \wedge \, \omega_2 \, .
\eea
Notice that
\bea
\frac{d \left(k(\tau) \, f(\tau)\right)}{d\tau} &=& \frac{1}{2} \, (\tau \, \coth\tau-1) 
\, \left( \coth\tau - \frac{\tau}{\sinh^2\tau} \right) \, . \label{derivativefk}
\eea
To calculate Hodge dual of the forms it is useful to have 
\bea
\sqrt{|g|} &=& \frac{\sqrt{h_0(\tau)} \, \epsilon^4 \, \sinh^2\tau}{96} \, ,
\eea
where $|g|$ is the determinant of the metric (\ref{metricwarped}).

Certain identities are useful for the explicit checks of the EOM. Indeed,
they are
\bea
2 \, \omega_2 \, \wedge \, \omega_2 &=& g^1 \, \wedge \, g^2 \, 
\wedge \, g^3 \, \wedge \, g^4 \, ,  \label{id}
\eea
\bea
dg^5 &=& -(g^1 \, \wedge \, g^4+g^3 \, \wedge \, g^2) \, ,  \label{id0}
\eea
\bea
dg^5 \, \wedge \, dg^5 &=& - 4 \, \omega_2 \, \wedge \, \omega_2 \, ,  \label{idi}
\eea
\bea
H_3 \, \wedge \, d\tau \, \wedge \, g^5 &=& 0 \, , \label{id1}
\eea
\bea
H_3 \, \wedge \, dg^5 &=& 0 \, . \label{id2}
\eea

~

\subsection*{Appendix C: Kuperstein-Sonnenschein explicit solutions}

~

Here we introduce the solutions of the ODE system of 
Eqs.(\ref{eqsODE}) and (\ref{eqYq}) obtained by Kuperstein and Sonnenschein.

First notice that solving the type IIB supergravity EOM, it is found
a first order ODE system for $F(\tau)$, $f(\tau)$ and $k(\tau)$, from where
the following solutions result \cite{Kuperstein:2003yt}
\bea
F(\tau) &=& \frac{1}{2} -  \frac{\tau}{2 \sinh\tau} + C_1 \, 
\left( \cosh\tau - \frac{\tau}{\sinh\tau} \right) + C_2 \,
\frac{1}{\sinh\tau} \, , \nonumber \\
f(\tau) &=& \frac{\tau \, \coth\tau - 1}{2 \sinh\tau} \, (\cosh\tau -1) + \nonumber \\
&& C_1 \, \left( 2 \tau - \sinh\tau - \tanh\frac{\tau}{2} - 
\frac{\tau}{2 \cosh^2\frac{\tau}{2}} \right) + C_2 \,
\frac{1}{2 \cosh^2\frac{\tau}{2}} + C_3 \, , \nonumber \\
k(\tau) &=& \frac{\tau \, \coth\tau - 1}{2 \sinh\tau} \, (\cosh\tau +1) + \nonumber \\
&& C_1 \, \left( 2 \tau + \sinh\tau - \coth\frac{\tau}{2} + 
\frac{\tau}{2 \sinh^2\frac{\tau}{2}} \right) - C_2 \,
\frac{1}{2 \sinh^2\frac{\tau}{2}} + C_3 \, . \label{Ffk}
\eea
Now, for Eqs.(\ref{eqsODE}) and (\ref{eqYq}) the solutions can be written as
\bea
\bar f(\tau) &=& \lambda_1(\tau) \, f_1(\tau) + \lambda_2(\tau) \, f_2(\tau) +
\lambda_3(\tau) \, f_3(\tau) \, , \\
\bar k(\tau) &=& \lambda_1(\tau) \, k_1(\tau) + \lambda_2(\tau) \, k_2(\tau) +
\lambda_3(\tau) \, k_3(\tau) \, , \\
\bar F(\tau) &=& \lambda_1(\tau) \, F_1(\tau) + \lambda_2(\tau) \, F_2(\tau) +
\lambda_3(\tau) \, F_3(\tau) \, ,
\eea
where $f_i(\tau)$, $k_i(\tau)$ and $F_i(\tau)$, for $i=1$, $2$ and $3$ are the 
functions in Eqs.(\ref{Ffk}) above that appear multiplied by the constants
$C_i$'s, respectively. The $\lambda_i$'s \cite{Kuperstein:2003yt} are
\bea
\lambda_1(\tau) &=& \frac{1}{2} \, \int_\tau^\infty \, 
\frac{k_0'(x)+f_0'(x)}{\sinh x} \, \bar y(x) \, dx \, , \\
\lambda_2(\tau) &=& \frac{1}{2} \, \int_0^\tau \, \left(
\frac{k_0'(x)+f_0'(x)}{2} \, \bar y(x) \, 
\left( \cosh x - \frac{x}{\sinh x} \right) -\frac{X}{P} \, h_0(x) \, \sinh^2x 
\right) dx \, , \\
\lambda_3(\tau) &=& - \int_\tau^\infty \, \left(
(f_0'(x)-k_0'(x)) \, \bar y(x) - \frac{1}{2} \sum_{i=1}^2
\left( (k_i(x)+f_i(x)) \lambda_i'(x) \right) + \frac{X}{P} \, h_0(x)  
\right) dx \, , \nonumber \\
\eea
while $\bar q$ is given by
\bea
\bar q(\tau) &=& \frac{1}{\rho(\tau)} \, \int_0^\tau \rho(x) \,
[\frac{2}{3} ({\dot {\bar Y}}-{\dot {\bar q}})+ \frac{1}{\sqrt{3}}
\, e^{-4p_0(x)-6q_0(x)} \, {\tilde {L}}_0(x)  \, \times \nonumber \\
&& \left( -4 \bar p(x)+\frac{1-F_0(x)}{L_0(x)} \, \bar f(x) +
\frac{F_0(x)}{L_0(x)} \, \bar k(x) + \frac{k_0(x)-f_0(x)}{L_0(x)} \, \bar F(x) \right)] \, dx \, ,
\eea
where 
\bea
\rho(\tau) &=& \exp{[2\sqrt{3} \int_{\tau_0}^\tau 
\, e^{-4p_0(x)-6q_0(x)} \, {\tilde {L}}_0(x) \, dx]} \, .
\eea

\end{document}